\documentclass[conference]{IEEEtran}
\IEEEoverridecommandlockouts

\usepackage{cite}
\usepackage{amsmath,amssymb,amsfonts}
\usepackage{algorithmic}
\usepackage{graphicx}
\usepackage{textcomp}
\usepackage{xcolor}
\usepackage{subfig}
\usepackage{braket}
\usepackage{multirow}

\def\BibTeX{{\rm B\kern-.05em{\sc i\kern-.025em b}\kern-.08em
    T\kern-.1667em\lower.7ex\hbox{E}\kern-.125emX}}

\definecolor{pink}{HTML}{F282B4}
\definecolor{corn}{HTML}{6495ED}

\begin{document}

\title{Quantum machine learning framework for longitudinal biomedical studies \\

}
\author{\IEEEauthorblockN{Maria Demidik}
\IEEEauthorblockA{
\textit{Deutsches Elektronen-Synchrotron DESY}\\
Zeuthen, Germany}
\IEEEauthorblockA{
\textit{The Cyprus Institute}\\
Nicosia, Cyprus \\
maria.demidik@desy.de}
\and
\IEEEauthorblockN{Filippo Utro}
\IEEEauthorblockA{
\textit{IBM Research}\\
Yorktown Heights,NY, USA \\
futro@us.ibm.com}
\and
\IEEEauthorblockN{Alexey Galda}
\IEEEauthorblockA{
\textit{Moderna}\\
Cambridge, MA, USA \\
alexey.galda@modernatx.com}
\and
\IEEEauthorblockN{Karl Jansen}
\IEEEauthorblockA{
\textit{The Cyprus Institute}\\
Nicosia, Cyprus}
\IEEEauthorblockA{
\textit{Deutsches Elektronen-Synchrotron DESY}\\
Zeuthen, Germany \\
karl.jansen@desy.de}
\and
\IEEEauthorblockN{Daniel Blankenberg}
\IEEEauthorblockA{\textit{Lerner Research Institute} \\
\textit{Cleveland Clinic}\\
Cleveland, Ohio, USA \\
blanked2@ccf.org}
\and
\IEEEauthorblockN{Laxmi Parida}
\IEEEauthorblockA{
\textit{IBM Research}\\
Yorktown Heights, NY, USA \\
parida@us.ibm.com}
}

\maketitle

\begin{abstract}
Longitudinal biomedical studies play a vital role in tracking disease progression, treatment response, and the emergence of resistance mechanisms, particularly in complex disorders such as cancer and neurodegenerative diseases. However, the high dimensionality of biological data, combined with the limited size of longitudinal cohorts, presents significant challenges for traditional machine learning approaches. In this work, we explore the potential of quantum machine learning (QML) for longitudinal biomarker discovery. We propose a novel modification to the instantaneous quantum polynomial time (IQP) feature map, designed to encode temporal dependencies across multiple time points in biomedical datasets. Through numerical simulations on both synthetic and real-world datasets—including studies on follicular lymphoma and Alzheimer’s disease—we demonstrate that our longitudinal IQP feature map improves the ability of quantum kernels to capture intra-subject temporal patterns, offering a promising direction for QML in clinical research.
\end{abstract}


\section{Introduction}

Longitudinal studies are essential in biomedical research for capturing dynamic changes in biological, physiological, and cognitive processes over time. Unlike cross-sectional studies that provide only a snapshot, longitudinal data enables tracing individual trajectories, identifying early markers of disease progression, and assessing treatment efficacy or resistance mechanisms~\cite{10.1158/1538-7445.AM2018-3001}. Such insights are particularly important in chronic and progressive diseases. For example, many cancer therapies initially suppress tumor growth but eventually fail due to the emergence of resistance mechanisms. In the case of follicular lymphoma (FL), transformation to a more aggressive disease state remains a critical clinical challenge~\cite{bai2024multi}. Similarly, Alzheimer's disease (AD) is characterized by a gradual neurodegenerative process, where patients may appear cognitively stable for extended periods before experiencing rapid decline~\cite{marcus_open_2010}.

Longitudinal data consists of repeated measurements of a set of features collected from the same subjects across time, enabling the study of intra-subject variability and temporal trends. Despite their clinical relevance, longitudinal biomedical datasets often suffer from small cohort sizes, irregular or sparse sampling intervals, and high acquisition costs, particularly when involving omics or imaging modalities. These factors present significant challenges for traditional machine learning methods, which typically require large amounts of labeled data to generalize effectively.

Quantum machine learning (QML) presents a promising alternative to classical approaches~\cite{biamonte2017quantum}, particularly in settings with limited sample sizes. Notably, QML models have been shown to generalize from fewer data points~\cite{caro_generalization_2022}, and to offer data-dependent predictive advantages through quantum kernels~\cite{huang_power_2021}. These properties make QML a compelling tool for longitudinal studies~\cite{flother_how_2024}, and have also been explored in broader biomedical and drug discovery contexts~\cite{smaldone2024quantum}.

However, most existing QML models are developed under the assumption that input data are independently and identically distributed (i.i.d.), an assumption that fails to account for the temporal dependencies inherent in longitudinal data. In biomedical applications, neglecting such structure can result in suboptimal representations and reduced predictive performance.

In this work, we propose a quantum kernel framework tailored for longitudinal biomedical studies. It has been shown that customizing the kernel to the target application can enhance the performance of QML models~\cite{glick_covariant_2024}. In quantum kernel methods, classical data are encoded into quantum states via feature maps, which define the structure of the kernel. One widely adopted feature map is the instantaneous quantum polynomial time (IQP) feature map~\cite{havlivcek2019supervised}. We introduce a longitudinal IQP feature map that explicitly incorporates temporal dependencies across multiple time points, thereby extending the representational capacity of quantum kernels for temporal biomedical data.

We evaluate the proposed feature map on synthetic data and two public longitudinal biomedical datasets. Our results show that the longitudinal IQP feature map improves the modeling of disease dynamics when compared to the IQP feature map.

The remainder of this work is structured as follows. Section~\ref{sec:framework} introduces the necessary background on kernel methods. Section~\ref{sec:data-encoding} presents the IQP and longitudinal IQP feature maps, along with numerical simulations that illustrate their ability to model temporal dependencies. Section~\ref{sec:num-res} describes the empirical evaluation on real-world biomedical datasets. Finally, Section~\ref{sec:discussion} discusses the implications of our findings for longitudinal biomarker discovery and broader biomedical applications.

\section{Framework}
\label{sec:framework}

Support vector machines (SVMs) provide a well-established framework for classification in supervised learning~\cite{svm}. By leveraging high-dimensional feature mappings, SVMs project input data into a transformed feature space, where an optimal hyperplane is constructed to separate distinct classes. This approach relies on the kernel trick, which enables efficient computation of inner products in the high-dimensional space via kernel functions~\cite{kernel_methods}.

Let $\boldsymbol{x}$ and $\boldsymbol{x}'$ be two real-valued input vectors of length $n$ ($n$ is the number of features), and let $\phi(\boldsymbol{x})$ denote the associated feature map. The kernel trick allows computing the inner product in the feature space without explicitly evaluating $\phi$, using a kernel function $K$ such that
\begin{equation}
K(\boldsymbol{x},\boldsymbol{x}') = \langle \phi(\boldsymbol{x}), \phi(\boldsymbol{x}') \rangle. 
\end{equation}
The choice of feature map $\phi$ (implicitly defined by $K$) plays a crucial role in the model’s ability to capture complex, nonlinear decision boundaries. A widely used example is the radial basis function (RBF) kernel, defined by
\begin{equation}
K(\boldsymbol{x},\boldsymbol{x}')  = \mathrm{exp}(-\gamma || \boldsymbol{x} - \boldsymbol{x}' ||^2)\,,
\end{equation}
where $\gamma > 0$ is a scale parameter controlling the kernel’s sensitivity to distance.

\begin{figure}[!h]
    \centering
    \includegraphics[width=0.7\columnwidth]{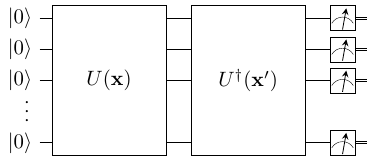}
    \caption{Quantum circuit used to evaluate the quantum kernel. The kernel $K(\boldsymbol{x},\boldsymbol{x}')$ is computed by measuring the squared overlap between two quantum states, as defined in Eq.~\eqref{fidelity-kernel}.}
    \label{fig:q-kernel}
\end{figure}

Quantum computers offer an alternative framework for defining feature maps through parametrized quantum circuits. Let $\ket{0}^{\otimes n}$ denote the $n$-qubit computational basis state. A parametrized unitary operation $U(\boldsymbol{x})$, acting on $n$ qubits, defines a quantum feature map by encoding the classical input $\boldsymbol{x}$ into a quantum state: $\ket{\psi(\boldsymbol{x})} = U(\boldsymbol{x}) \ket{0}^{\otimes n}$. The corresponding kernel function $K$ is then constructed as the squared overlap (fidelity) between quantum states,
\begin{equation}
K(\boldsymbol{x},\boldsymbol{x}')  = \left| \langle \psi(\boldsymbol{x}) | \psi(\boldsymbol{x}') \rangle  \right|^2\,,
\label{fidelity-kernel}
\end{equation}
which can be evaluated using the quantum circuit illustrated in Figure~\ref{fig:q-kernel}.

In this study, we employ a quantum feature map based on instantaneous quantum polynomial time (IQP) circuits~\cite{iqp_circuit}, a special class of quantum circuits that play a significant role in quantum complexity theory and have been widely used as kernels in quantum machine learning~\cite{havlivcek2019supervised,PhysRevA.106.042407}. 

\begin{figure*}[!t]
    \centering
    \includegraphics[width=0.8\linewidth]{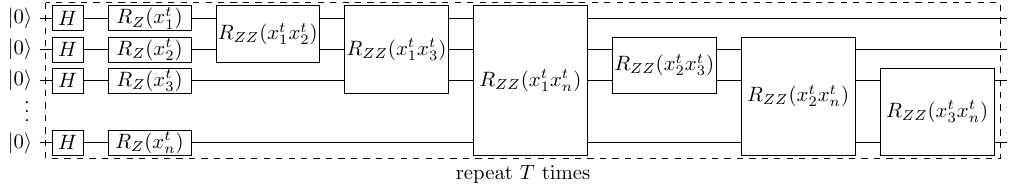}
    \caption{IQP circuit used to define the quantum feature map for longitudinal data. The input consists of $n$ features measured over $T$ time points, representing longitudinal observations. For each time point $t \in \{1, \dots, T\}$, a separate IQP layer is applied to encode the corresponding feature vector $\boldsymbol{x}^t$. Each layer consists of Hadamard gates followed by a diagonal unitary $U_D(\boldsymbol{x}^t)$, composed of single- and two-qubit rotations such as $R_Z$ and $R_{ZZ}$}
    \label{fig:iqp-fm}
\end{figure*}

\section{Longitudinal data embedding}
\label{sec:data-encoding}

Formally, an IQP feature map encodes the input $\boldsymbol{x}$ of length $n$ into a quantum state as
\begin{equation}
\ket{\psi(\boldsymbol{x})} = U_D(\boldsymbol{x}) H^{\otimes n} \ket{0}^{\otimes n},
\end{equation}
where $U_D$ is a diagonal unitary operator, typically composed of parametrized single-qubit and two-qubit rotations such as $R_Z$ and $R_{ZZ}$. 

The most straightforward way to encode temporal data via IQP feature map is to encode each time point independently as layers of the feature map. Then, the encoding of the input $\boldsymbol{x}$ characterized by $T$ time points into a quantum state is given by
\begin{equation}
    |\psi(x)\rangle = U_D\left( \boldsymbol{x}^T\right) H^{\otimes n} \cdots U_D\left( \boldsymbol{x}^1\right)H^{\otimes n}|0\rangle^{\otimes n}\,,
\end{equation}
where each $U_D\left( x^t \right)$ is parametrized by the input at time point $t$. An illustration of the IQP feature map that encodes longitudinal data is provided in Figure~\ref{fig:iqp-fm}.

\begin{figure*}[!t]
    \centering
    \includegraphics[width=\linewidth]{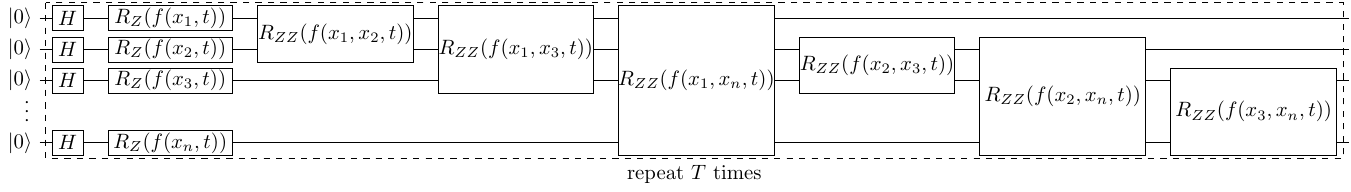}
    \caption{Longitudinal IQP circuit used to encode time-dependent feature interactions. Unlike the IQP feature map, which applies a separate layer for each time point, the longitudinal IQP circuit encodes data through time-dependent functions evaluated at each layer. Specifically, the quantum circuit parameters are modulated by functions of the form $f(x_i, t)$ and $f(x_i, x_j, t)$, as defined in Eqs.~\eqref{eq:fn-single} and \eqref{eq:fn-double}, respectively. These functions capture both temporal dynamics and pairwise feature interactions, enabling a richer embedding of longitudinal data into the quantum state space.}
    \label{fig:iqp-longit-fm}
\end{figure*}

While this approach enables encoding temporally structured inputs using the IQP feature map, it treats each time point as an independent layer, thus failing to capture correlations across time within a single sample. This may lead to the lack of intra-sample temporal correlation modeling. To address this limitation and account for temporal dependency within a sample, we propose a modification to the IQP feature map by altering how input features are used to parameterize quantum gates. Rather than treating each time point independently, we define a functional dependency over time in the feature encoding. Specifically, for layer $t$ (corresponding to the $t$-th time point), the input parameters for the single-qubit gates are obtained as follows:
\begin{equation}
    f(x_i, t) = \frac{1}{t} \sum_{m=1}^t x_i^m\,,
\label{eq:fn-single}
\end{equation}
where $x_i^m$ denotes the value of feature $i$ at time point $m$. Correspondingly, the parameters of the two-qubit gates at layer $t$ are then given by:
\begin{equation}
    f(x_i, x_j, t) = f(x_i, t) \cdot f(x_j, t).
\label{eq:fn-double}
\end{equation}
This cumulative form introduces a temporal smoothing effect, enabling the kernel function to capture progressive intra-subject changes over time. The IQP feature map with modified encoding is referred to as the longitudinal IQP feature map, and it is illustrated in Figure~\ref{fig:iqp-longit-fm}. 

We evaluate the IQP and longitudinal IQP feature maps using a synthetic dataset. The generated dataset consists of 100{,}000 samples, each described by a single feature that takes values in the range [0, $2\pi$]. To simulate temporal structure, each sample includes two time points, and we model dependencies by computing the fidelity between each sample and a fixed reference sample with both time point values set to $\pi/2$.

\begin{figure}[ht]
\centering
\subfloat[]{
\includegraphics[width=0.406\columnwidth]{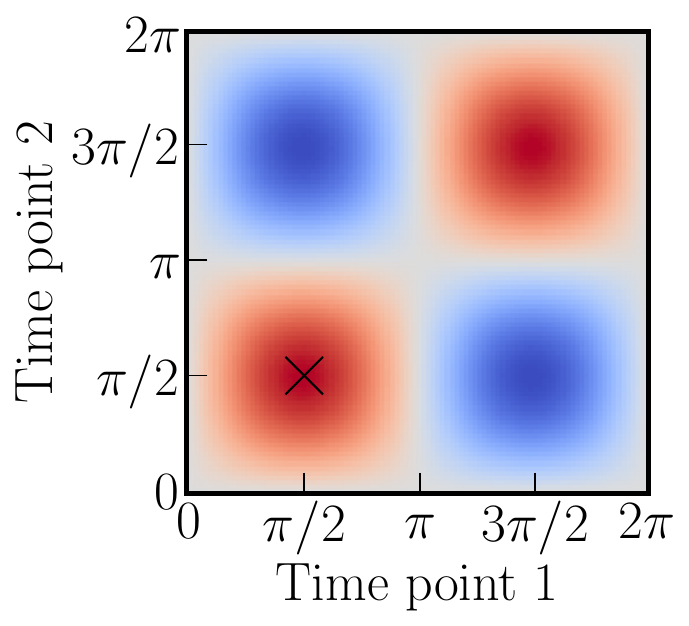}
\label{fig:iqp-nonmodified}
}
\subfloat[]{
\includegraphics[width=0.47\columnwidth]{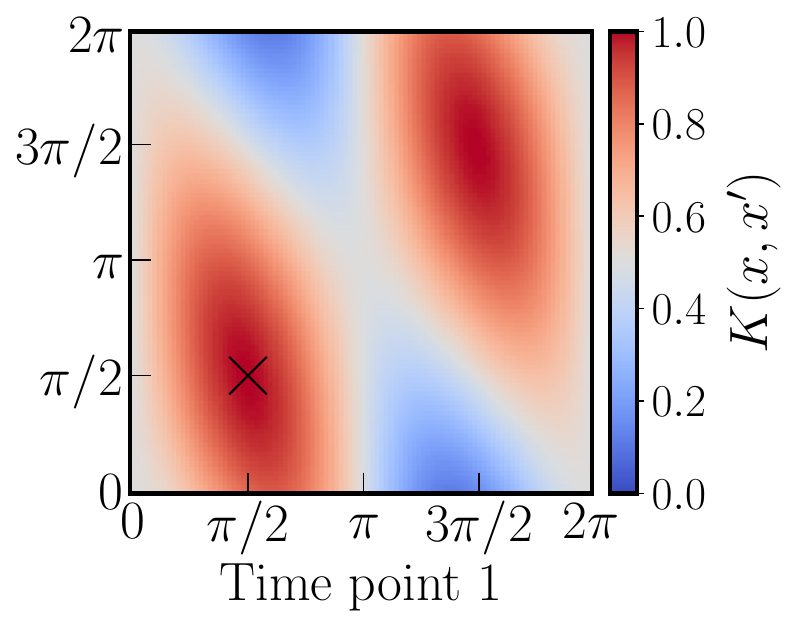}
\label{fig:iqp-longitudinal}
}
\caption{Comparison of the IQP and longitudinal IQP kernels for a two-time-point single-feature setting. The kernel is computed with respect to a fixed reference point $(\pi/2, \pi/2)$, and evaluated across all possible values of $x^1, x^2 \in [0, 2\pi]$, corresponding to time points 1 and 2, respectively. The left plot (a) shows the kernel values obtained using the IQP feature map, while the right plot (b) depicts the results from the longitudinal IQP feature map. The reference point $(\pi/2, \pi/2)$ is marked with an `X' in both plots.}
\label{fig:iqp-results}
\end{figure}

As shown in Figure~\ref{fig:iqp-nonmodified}, for the IQP feature map, the fidelity between generated samples and the stationary sample varies periodically and symmetrically with changes in the second time point, when the first time point is fixed. This loss of temporal dependency within a sample is undesirable for biomedical applications. Thus, leveraging the IQP feature map in longitudinal studies may result in limited performance of a QML model.

In contrast, in Figure~\ref{fig:iqp-longitudinal}, the longitudinal IQP feature map enables the kernel function to capture changes in the second time point as a function of the first time point. This behavior indicates the potential of the longitudinal IQP feature map to effectively model temporal patterns, which is particularly relevant for biomedical studies involving disease progression or treatment response.

\section{Numerical results}
\label{sec:num-res}

To evaluate our approach we use two public biomedical datasets associated with following diseases:

\paragraph{Follicular Lymphoma (FL)} is the most common indolent type of B-cell non-Hodgkin lymphoma. In Ref.~\cite{Bai2024-cl}, the authors perform a multi-omics analysis to study early relapse and transformation of FL to aggressive disease, associated with inferior outcome. Despite identifying novel biomarkers in FL pathogenesis, it remains challenging to identify subjects that will undergo mutations in the suggested biomarkers. Based on the available RNA-seq data from Ref.~\cite{Bai2024-cl}, we consider the task of classifying 32 subjects into two groups: relapsed non-transformed FL (nFL) and transformed FL (tFL). Each subject is characterized by multiple biopsies, ranging in the amount from one to three. We select first 20 significantly differentially expressed genes leveraging Welch's t-test prior to running kernel methods.

\paragraph{Alzheimer's disease (AD)} is a progressive neurodegenerative disorder. The data was collected by the Open Access Series of Imaging Studies (OASIS)~\cite{marcus_open_2010}. The data of 150 subjects is present with a longitudinal collection of 3 or 4 individual MRI scans. In addition, the subjects' neurodegenerative status is described by clinical dementia rating (CDR), normalized whole-brain volume (nWBV), estimated total intracranial volume (eTIV), mini-mental state examination score (MMSE) and atlas scaling factor (ASF). Initially 64 subjects were diagnosed as demented, while 72 remained in the non-demented group throughout the study. Another 14 subjects transitioned in their status from non-demented to demented. Following the approach of Ref.~\cite{BATTINENI2019100200}, we exploit the described features of neurodegenerative status to perform binary classification of subjects based on the CDR status. In future, the work can be extended by incorporating MRI scan data into the framework.

We evaluate the performance of three feature maps: classical RBF, IQP, and longitudinal IQP for the two biomedical longitudinal datasets. To apply the RBF kernel to longitudinal data, we provide all the time points as separate features. Using these feature maps, we train SVM models via standard convex optimization techniques~\cite{BATTINENI2019100200}. We report the train and test accuracy of binary classification in Table~\ref{tab:results}. 

\begin{table}[htbp]
\centering
\caption{Classification performance across datasets using different feature maps. The table reports training and test accuracies for two datasets—Follicular lymphoma and Alzheimer’s disease—using three types of feature maps: classical RBF, IQP, and longitudinal IQP. }
\label{tab:results}
\begin{tabular}{cccc}
\textbf{Dataset} & \textbf{Feature map} & \textbf{Train accuracy} & \textbf{Test accuracy}\\
\hline
\hline
\multirow{3}{*}{\shortstack{Follicular\\lymphoma}}
  & RBF & 25\,/\,25 & 6\,/\,7 \\
  & IQP & 25\,/\,25 & 6\,/\,7\\
  & Longitudinal IQP & 25\,/\,25 & 7\,/\,7 \\
\hline
\multirow{3}{*}{\shortstack{Alzheimer's\\disease}}
  & RBF & 120\,/\,120 & 19\,/\,30\\
  & IQP & 111\,/\,120 & 18\,/\,30 \\
  & Longitudinal IQP & 110\,/\,120 & 22\,/\,30 \\
\hline
\end{tabular}
\end{table}

While all models achieve perfect or near-perfect training accuracy, the longitudinal IQP feature map consistently demonstrates improved test performance relative to the IQP approach. This suggests enhanced generalization capability, particularly when modeling data with an underlying temporal or progressive structure. Across both datasets, the results indicate that incorporating temporal dependencies via the longitudinal IQP feature map can be beneficial for analyzing diseases with progressive dynamics, such as AD and FL.

\section{Discussion}
\label{sec:discussion}

In this work, we introduced the longitudinal IQP feature map designed to encode temporal dependencies inherent in longitudinal biomedical datasets. The proposed feature map explicitly incorporates temporal correlations across time points within a sample. The longitudinal IQP feature map is based on the IQP feature map that uses a cumulative encoding scheme: each circuit layer aggregates information from the current and all previous time points. This enables quantum kernels to capture intra-subject temporal correlations, which are essential for understanding progressive disease dynamics.

Numerical experiments on biomedical datasets demonstrate that the longitudinal IQP feature map enhances kernel expressivity for temporally structured data. In particular, we observed that, while the RBF kernel and the kernel based on the IQP feature map achieved high training accuracy, leveraging the longitudinal IQP feature map consistently resulted in higher test performance. These findings underscore the value of adapting quantum feature maps to reflect the structure of the input data, especially in contexts where modeling time progression is critical.

Longitudinal biomedical datasets often consist of small cohorts. Although QML models hold promise for generalization properties in low-data regimes, it remains challenging to find a public dataset with big enough cohort size to evaluate the generalization capabilities of the proposed model. Future work should assess the robustness and scalability of the longitudinal IQP feature map on larger cohorts with complex temporal patterns. Additionally, extending this approach to multimodal data, e.g. incorporating longitudinal imaging alongside molecular profiles, could further enhance its utility in biomarker discovery for neurodegenerative and oncological diseases. Exploring alternative QML architectures, such as quantum reservoir computing~\cite{suzuki_natural_2022} or quantum recurrent models~\cite{bausch_recurrent_2020}, may also provide complementary advantages for sequential biomedical tasks.

Finally, we note that the results for all feature maps, including RBF, may benefit from additional hyperparameter tuning. However, the primary goal of this work is to emphasize the importance of adapting QML models to domain-specific data structures. Even if near-term quantum models remain classically simulable, their design and evaluation offer a novel computational perspective for modeling disease dynamics and advancing data-driven clinical research.

\section*{Acknowledgment}
This work is supported with funds from the Ministry of Science, Research and Culture of the State of Brandenburg within the Center for Quantum Technologies and Applications (CQTA). This work is funded within the framework of QUEST by the European Union’s Horizon Europe Framework Programme (HORIZON) under the ERA Chair scheme with grant agreement No.\ 101087126.

\bibliographystyle{unsrt}  
\bibliography{refs}
\end{document}